# Mining Public Transit Ridership Flow and Origin-Destination Information from Wi-Fi and Bluetooth Sensing Data

Ziyuan Pu[a], Zhiyong Cui[a], Meixin Zhu[a], Yinhai Wang[a]∗

*[a] Deparment of Civil and Environmental Engineering, University of Washington, Seattle, WA, US*

**Abstract**

Transit ridership flow and origin-destination (O-D) information is essential for enhancing transit network design, optimizing transit route and improving service. The effectiveness and preciseness of the traditional survey-based and smart card data-driven method for O-D information inference have multiple disadvantages due to the insufficient sample, the high time and energy cost, and the lack of inferring results validation. By considering the ubiquity of smart mobile devices in the world, several methods were developed for estimating the transit ridership flow from Wi-Fi and Bluetooth sensing data by filtering out the non-passenger MAC addresses based on the predefined thresholds. However, the accuracy of the filtering methods is still questionable for the indeterminate threshold values and the lack of quantitative results validation. By combining the consideration of the assumed overlapped feature space of passenger and non-passenger with the above concerns, a three steps data-driven method for estimating transit ridership flow and O-D information from Wi-Fi and Bluetooth sensing data is proposed in this paper, three steps including extracting the features of the representative characteristics of each detected Wi-Fi and Bluetooth MAC address, clustering the detected MAC addresses into passenger and non-passenger clusters using Fuzzy C-Means (FCM) clustering algorithm, and estimating the population ridership flow and O-D information using Random Forest (RF) regression based on the number of MAC address in passenger cluster. The observed ridership flow is used as ground truth for calculating the performance measurements. According to the results, the proposed approach outperformed all selected baseline models and existing filtering methods. Based on the proposed approach, a real-time transit ridership flow and O-D information monitoring system is designed. The findings of this study can help to provide real-time and precise transit ridership flow and O-D information for supporting transit vehicle management and the quality of service enhancement.

## 1. Introduction

Real-time public transit ridership flow and origin-destination (O-D) information are crucial for transit network planning and routes optimization. Traveler surveys have been employed for acquiring such information in the majority of previous relevant research (Ben-Akiva et al., 1985). However, the effectiveness and efficiency of survey-based methods are questionable in terms of high time and energy cost, and biased statistics (Wardman, 1988). As the smart card becoming widely-used, some researchers have developed methods for estimating ridership and O-D related information based on transit smart card data (Kieu et al., 2015; Kusakabe and Asakura, 2014; Ma et al., 2017, 2013, 2012; Morency et al., 2007). Nevertheless, the lack of quantitative validation makes the accuracy of smart card data-based methods still doubtful. Nowadays, it is stated that more than 80% of individuals carried at least one Wi-Fi and Bluetooth mobile device in daily life ("• Global mobile phone internet user penetration 2019 | Statistic," n.d., "• Smartphone penetration in the US (share of population) 2010-2021 | Statistic," n.d.). Thus, estimating transit ridership and O-D information based on Wi-Fi and Bluetooth sensing data has the great potential to be a more reliable method.

The mechanism of Wi-Fi and Bluetooth sensing technology is capturing the hardware Media Access Control (MAC) address of the discoverable mobile devices through Wi-Fi management frame and Bluetooth slave response message within a specific detection range (Malinovskiy et al., 2010). If the Wi-Fi or Bluetooth communication function of a mobile device is turned on and the mobile device is not paired with an access point or other devices through Wi-Fi or Bluetooth protocol, such mobile device is discoverable. Since the hardware MAC address is a globally unique identifier, it is easy to monitor the boarding and alighting information of each passenger by sensing the existence of the passenger's mobile devices. However, since the detection range of Wi-Fi and Bluetooth sensing usually is larger than the inside space of transit vehicles, the mobile devices carried by non-passengers are still possible to be detected. Thus, separating the mobile devices belonging to passengers and non-passengers is crucial for estimating ridership and O-D information from Wi-Fi and Bluetooth sensing data. Previously, several studies shed light on solving this problem based on filtering methods (Dunlap et al., 2016; Mishalani et al., 2016). Basically, several empirically preset thresholds were used to filter out the data potentially belonging to non-passengers in terms of the time duration of MAC address existence, the number of times detected, the distances between the mobile devices and the transit stations for the first and the last detection, etc. However, the results of majority studies are barely convinced due to the lack of quantitative performance measurements. Since the true O-D information collection is costly and labor intensive, only a few studies provided the comparison of observed ridership flow and the estimated ridership flow, including the number of onboard, boarding and alighting passenger of each stop, with the estimated ridership flow based on filtering results (Hidayat et al., 2018; Ji et al., 2017; Oransirikul et al., 2014). The obvious gaps between the observed data and the filtering results indicated the considerable errors caused by such hard threshold filtering methods. Hence, an accurate and effective method for separating the MAC address data belonging to passengers and non-passengers is quite needful.

---

∗ Corresponding author.
*E-mail address:* yinhai@uw.edu.

The main disadvantage of the filtering method is that the threshold value is hard to be determined. The potential reason could be the features of MAC address data of passenger and non-passenger could be overlapped. For example, a transit vehicle travels with another vehicle side by side for a distance, the features of the mobile devices carried by onboard passengers could be similar to the features of the mobile devices in the other vehicle. Based on the consideration of the overlapped feature spaces of passengers and non-passengers, a Fuzzy C-Means (FCM) clustering algorithm was proposed for separating passenger and non-passenger MAC addresses. Unlike hard or crisp clustering algorithms, e.g. K-Means clustering, FCM allows objects to have the possibility for belonging to all groups with a certain degree of membership. FCM is one of the most popular fuzzy-based clustering algorithms which is suitable for separating the clusters with ambiguous boundaries (Bezdek et al., 1984). Previously, FCM was used to deal with the not well-separated clusters in several scenarios in intelligent transportation engineering area, e.g. ship trajectories clustering (Gan et al., 2018), traffic volume based road groups clustering (Gastaldi et al., 2014).

Since the passengers detected by Wi-Fi and Bluetooth sensing is a limited sample of the population, a method targeting on estimating the population ridership based on the number of MAC addresses belonging to transit passengers is still needed. Previously, several methods were implemented to estimate the population based on the number of detected Wi-Fi and Bluetooth MAC addresses, including scaling with a fixed number (Duives et al., 2018), linear regression (Kostakos et al., 2010), power function and Fourier function based methods (Lesani and Miranda-Moreno, 2016). However, it is hard to claim which method is more appropriate without any quantitative measurements of the estimation errors. Among the existing methods, Asad and Miranda-Moreno (2016) conducted a performance comparison of power function and Fourier function for estimating the population number of pedestrians based on the detected Wi-Fi and Bluetooth MAC addresses. The proposed power function achieved a relatively high R-square value than the Fourier function. In addition, the R-square value of the proposed power function is also much higher than linear regression method in other studies (Kostakos et al., 2013), which could be an indicator of the non-linear relationship between the population and the number of detected MAC addresses. Thus, to handle non-linearity among data, a Random Forest regression (RF) model (Breiman, 2001) is proposed for estimating the population ridership flow in this study, including the number of onboard, boarding and alighting passenger. For the performance comparison purpose, linear regression was selected as the baseline model to indirectly demonstrate the suitableness of the RF model.

Generally, the primary objective of this study is to develop a three-step data-driven methodology for estimating the transit ridership flow, including the number of onboard, boarding and alighting passenger, and O-D information from Wi-Fi and Bluetooth sensing data. More specifically, the objective can be split into four parts: 1), extracting multiple features for representing the characteristics of each MAC address; 2), proposing the FCM clustering algorithm for separating the passenger MAC addresses and non-passenger MAC addresses based on the extracted features; 3), estimating the population number of ridership flow based on the number of MAC addresses in the passenger cluster using the RF regression model; 4), designing a real-time transit ridership flow and O-D information monitoring system based on the proposed algorithm framework.

The remainder of the paper is organized as follows. The data collection tool and the data set used in this study are introduced at first. Furthermore, the features for representing the characteristics of each detected MAC address are proposed. Then, the method for passenger and non-passenger MAC address separation, the method for estimating the population number of ridership flow, and the performance evaluation measurements are introduced, respectively. After the explanation of the numerical results and the introduction of the passenger real-time monitoring system, the paper is summarized by concluding the research findings and the directions of future research.

## 2. Data Collection

The data used in this study were collected from 9 trips of 3 routes in Seattle. The detailed description of the data collection sensor, the study area, and the statistical description of the data set are introduced in the following sections.

### 2.1. Data Collection Sensor

The data collection sensor was developed based on Raspberry Pi. Raspberry Pi is a single-board computer developed in the United Kingdom. Raspberry Pi 3 Model B+ was employed in this study, which is the latest version of a series of products with 1.4 GHz 64-bit Quad-Core Processor and 1 GB RAM. The detailed technical parameters are referred from the official introduction material (Upton and Halfacree, 2014). Figure 1 shows the developed sensor used in this study. Raspberry Pi 3 Model B+, Wi-Fi adapter, Bluetooth adapter, GPS module, real-time clock module, and portable power bank were integrated. The following paragraphs describe each component in detail.

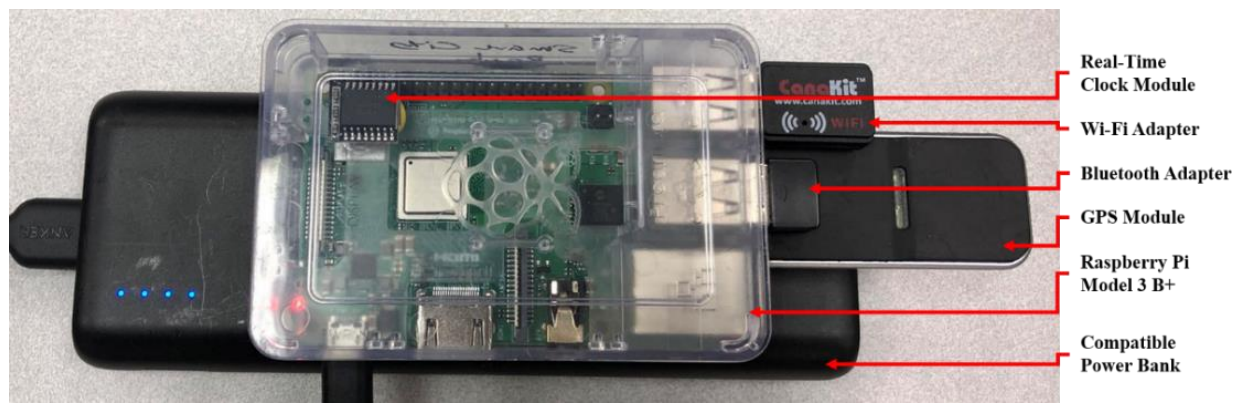

Figure 1: Data collection sensor.

To capture the MAC address of Wi-Fi management frames, the Wi-Fi 801.22ngb adapter was set in monitor mode in which the Wi-Fi adapter can capture the management frames sent from the discoverable mobile devices in Wi-Fi network within the detection range (Cunche, 2014). In this study, a Wi-Fi adapter with Ralink 5370 Wi-Fi chipset was used. Its detection range is about 200ft and its frequency range is 2.4 - 2.4835 GHz. For sensing the MAC address in Bluetooth slave response messages, the Bluetooth adapter was set to keep sending out inquiry request so that the tool can receive response messages of the discoverable mobile devices within the detection range. The Bluetooth adapter employed Bluetooth 4.0 BCM20702 chipset was used in this study. The detection range is about 60ft. For Wi-Fi and Bluetooth sensing data, each data point contains MAC address, timestamp, and received signal strength indicator (RSSI). In order to record the moving features of the data collection tool, a GPS module was employed to record the high-resolution latitude and longitude. As shown in Figure 1, the GPS module connected with Raspberry Pi through the USB port. The module employed a U-blox 7020 chipset with -162 dBm tracking sensitivity. The GPS module stored one data point half second, and each data point contains the current latitude, longitude, and timestamp.

In this study, data collection programs were running on the Raspberry Pi through a start-up script. Wi-Fi, Bluetooth, and GPS data were collected in parallel. The data integration of these three data sources is according to the timestamp. Raspberry Pi 3 has a real-time clock so that once the power is off the clock of Raspberry Pi would stop. If no internet connection or manual time synchronization service employed, the clock will not be synchronized with the current time. Even the GPS module can be used for time synchronization, it still has the possibility to fail in time correction if the GPS module has signal related issues. Thus, in order to avoid the issues caused by time, the DS 3231 RTC real-time clock module was employed. Then, the Raspberry Pi can synchronize time with the real-time clock module after booting. In addition, a portable power bank was used to supply power for the sensor's running.

## 2.2. Study Area

The study area of this research is three transit routes in the north of King County, including route 32, route 67 and route 372. Figure 2 shows the three routes on the map. Route 67 depicted in blue runs from University District to Northgate Transit Center, route 32 highlighted in red operates from Queen Avenue to Sand Point Way, and route 372 provides service along the route from Bothell to University District which is presented in green.

The data were collected from three trips of each route by the data collection sensor introduced in the last section. In each trip, the sensor was carried by volunteer seating in the middle of the vehicle. The sensor was powered on once the volunteer gets seated, and the sensor was powered off once the vehicle arrived at the last stop or the volunteer took off the vehicle. Since the MAC address data and GPS data were collected separately, a GPS location was assigned to each MAC address data point based on the timestamp. Figure 2 also visualizes all MAC address data by each route. The statistical description of the dataset is introduced in the next section.

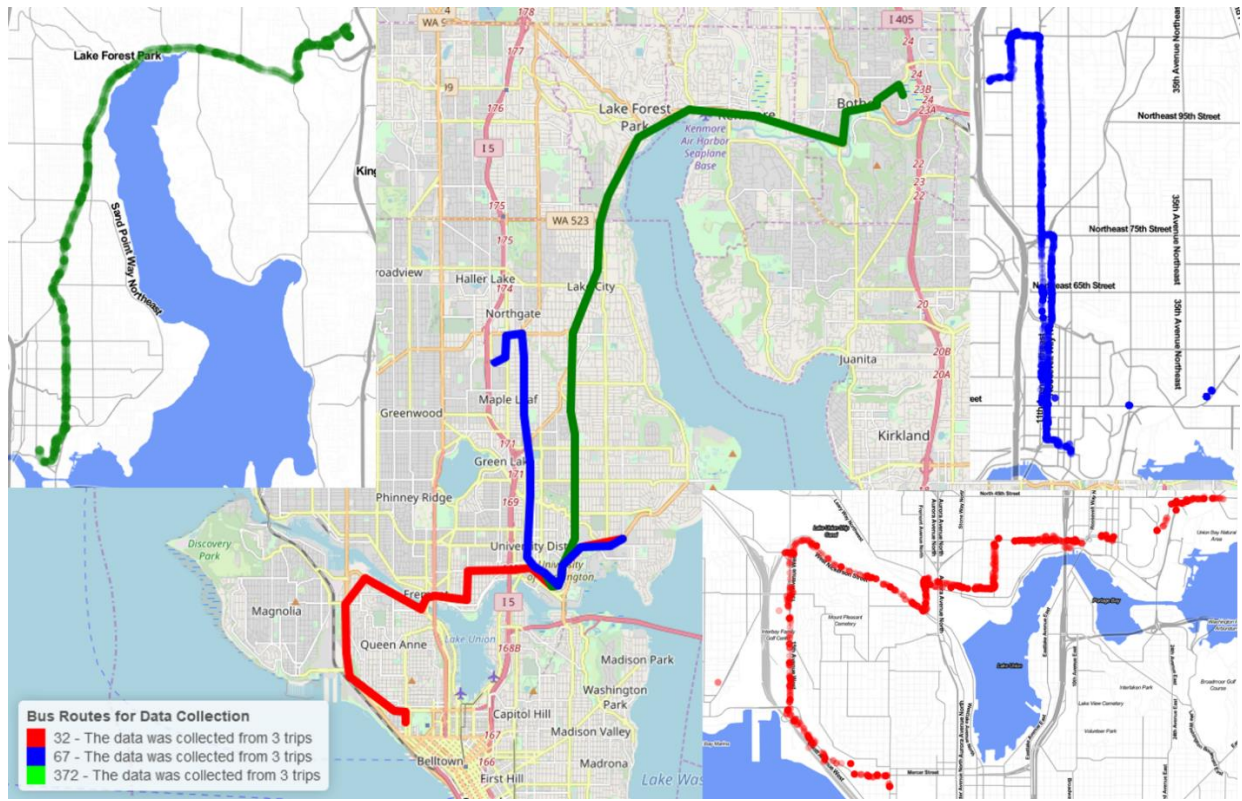

Figure 2: Study area.

## 2.3. Dataset Description

Table 1 shows the statistical description of each trip. There are 9 trips in total were traveled by the volunteer for collecting data. Notably, the number of stops is different from trip to trip. Since the vehicles only stop at the stations with waiting passengers or have passengers on vehicle requesting for taking off, only the stations where the vehicle has stopped were counted as stops in the dataset. In addition, only a part of the route was traveled for several trips, since the volunteer took off the vehicle before the vehicle arrived at the last station.

Besides the trip information, the amount of data point and the amount of unique MAC address collected from each trip also are provided. Based on the statistical description, 17806 MAC address data points were collected, including 16027 Wi-Fi data points and 1779 Bluetooth data points. The huge difference between the amount of Wi-Fi and Bluetooth data points is caused by the amount of discoverable Wi-Fi and Bluetooth devices and the different time intervals between the frame transmission of Wi-Fi and Bluetooth protocols. Totally, 5064 unique MAC addresses were detected, including 4859 via Wi-Fi network and 205 via Bluetooth network. Based on the dataset, averagely, one unique Wi-Fi MAC address would be collected out of 4 Wi-Fi data points and one unique Bluetooth MAC address out of 10 Bluetooth data points.

Table 1: Statistical description of the dataset

| Routes No. | Trip Date | Trip Start Time | Trip End Time | The Number of Stops | Number of Data Points | | Number of Unique MAC | |
|---|---|---|---|---|---|---|---|---|
| | | | | | Wi-Fi | Bluetooth | Wi-Fi | Bluetooth |
| 372 | 3/6/2018 | 7:35:00 | 8:32:00 | 21 | 2550 | 344 | 431 | 29 |
| | 3/6/2018 | 10:51:00 | 11:49:00 | 24 | 2055 | 344 | 854 | 53 |
| | 3/1/2018 | 11:03:00 | 11:51:00 | 28 | 3547 | 346 | 819 | 21 |
| 32 | 11/4/2018 | 16:55:37 | 17:21:57 | 12 | 904 | 172 | 294 | 8 |
| | 11/9/2018 | 18:40:49 | 19:26:26 | 24 | 2166 | 152 | 815 | 29 |
| | 11/9/2018 | 19:38:58 | 20:05:48 | 15 | 918 | 86 | 165 | 13 |
| 67 | 11/4/2018 | 15:05:15 | 15:47:26 | 27 | 1879 | 122 | 747 | 20 |
| | 11/8/2018 | 15:05:19 | 15:33:50 | 21 | 1351 | 88 | 555 | 18 |
| | 11/8/2018 | 15:38:10 | 16:04:44 | 19 | 657 | 125 | 179 | 14 |

*2.4. Feature Extraction*

As discussed in the introduction, the MAC addresses which belong to the mobile devices carried by other vehicles or pedestrians still have the possibility to be detected. In the following situations, the MAC addresses belonging to the mobile devices carried by non-passengers could be detected by the on-vehicle sensor.

- Fixed Wi-Fi or Bluetooth devices in roadside buildings within the detection range
- Mobile devices carried by people who are waiting for other transit vehicles at stations
- Mobile devices carried by pedestrian or bicyclist within the detection range
- Mobile devices in other vehicles within the detection range

For the fixed devices and devices carried by people at stations, the features of the MAC address data are quite different with the features of passenger MAC address. Intuitively, the MAC address should be detected very few times or in a very short time period. For the mobile devices in other vehicle or carried by pedestrian and bicyclist, even they travel parallel with the transit vehicle, the MAC features also would be different with MAC addresses which belong to passengers' mobile devices, e.g. the location of the first and the last detection could be far away from the nearest stations.

In order to depict the features of each MAC address, 9 features were extracted from the MAC address sensing data and GPS data for each unique MAC address, which are presented in Table 2. The features were categorized into two part in terms of which raw data set was used for feature extraction. MAC address features were extracted from Wi-Fi and Bluetooth sensing data, including the number of times detected, the detection duration, the average RSSI, and the maximum RSSI. Travel distance, average speed, maximum speed, and the distances of a vehicle to the nearest station when the MAC address was first and last detected were the five features which describe the moving status of the vehicle during the time period of each unique MAC address was detected. Based on such features calculated from MAC address data and GPS data, the MAC addressed belonging to passenger and non-passenger would be separated by the proposed methodology.

Table 2: Extracted features for characterizing each unique MAC address

| Categories | Features | Definition |
|---|---|---|
| MAC Address Features | Number of Times Detected | The number of times a unique MAC address is detected - (Times) |
| | Detection Duration | The total amount of time for a unique mac to be detected - (Seconds) |
| | Average RSSI | The average value of received signal strength indicator of given MAC - (dBm) |
| | Maximum RSSI | The maximum value of received signal strength indicator of given MAC - (dBm) |
| Vehicle Status | Least Distance Start | The distance between the location of vehicle and the nearest station when MAC address was first detected - (Meters) |
| | Least Distance End | The distance between the location of vehicle and the nearest station when the MAC address was last detected - (Meters) |
| | Travel Distance | The total travel distance of the vehicle between the first and the last detection of a unique MAC address - (Meters) |
| | Average Speed | The average speed of the vehicle between the first and the last detection of a unique MAC address - (Meters/Second) |
| | Maximum Speed | The largest speed of the vehicle between the first and the last detection of a unique MAC address - (Meters/Second) |

## 3. Methodology

In order to estimate the true number of onboard, boarding and alighting passengers of each stop based on the extracted features, two methods are required to be conducted continuously, including passengers' and non-passengers' MAC address separation, and estimating the population number of onboard, boarding and alighting passengers of each stop based on the number of clustered passengers' MAC addresses. The following section will introduce each method in detail and the validation methods and measurements as well.

### 3.1. Passenger and Non-Passenger MAC Address Separation using Fuzzy C-Means Clustering

As discussed in the section of the introduction, FCM clustering would be employed as the proposed methodology for passengers' and non-passengers' MAC address separation. Other than hard or crisp clustering algorithms, the fuzzy clustering algorithm assigns a certain degree of membership to a data point for all clusters, which indicates the data point can belong to any cluster (Zadeh, 1965). Thus, fuzzy clustering algorithms usually are useful when the boundaries among clusters are ambiguous(Xu and Wunsch, 2005), which satisfies the characteristics of the overlapped feature spaces of passengers and non-passengers.

FCM is one of the most popular fuzzy clustering algorithms. It attempts to minimize the cost function $J$ in Equation (1) which is the summation of the membership function of each data point. The membership function only depends on the distance to the center of each cluster. Then, assign each data point to the closest cluster in terms of the membership function. Let $\chi = (X_1, X_2, X_3, \cdots, X_N)$ denotes a set of $N$ MAC address data points to be partitioned into $C$ clusters which is 2 in this study, representing the clusters of passengers and non-passengers. $X_j = (x_1, x_2, x_3, \cdots, x_n)$ denotes $n$ features of each MAC address data points. Then, the cost function $J$ would be calculated as the following equation:

$$J = \sum_{j=1}^{N} \sum_{i=1}^{C} u_{ij}^m \|X_j - v_i\|^2 \qquad (1)$$

where $m$ is the parameter for controlling the fuzzification which is usually set to 2 (Hathaway and Bezdek, 2001). $u_{ij} \in [0,1]$ is the membership function of $j$th data point in cluster $i$, which represents the possibility that $j$th MAC address data point whether belongs to a passenger or not, thus, $\sum_{i=1}^{C} u_{ij}$ $(j = 1,2,\cdots,N)$. $v_i$ is the center of $i$th cluster, and $\|*\|$ is the similarity function of data point $X_j$ and the cluster center $v_i$. The cost function $J$ is minimized when the data points which are closer to the center of their clusters are assigned with higher membership values than the data points far from the centroid. The membership function and the center of each cluster are updated iteratively based on the following equations:

$$u_{ij} = \frac{1}{\sum_{k=1}^{c} \left(\frac{\|X_j - v_i\|}{\|X_j - v_k\|}\right)^{2/(m-1)}} \quad (2)$$

$$v_i = \frac{\sum_{j=1}^{N} u_{ij}^m X_j}{\sum_{j=1}^{N} u_{ij}^m} \quad (3)$$

Initially, the centers of each cluster were randomly selected from the data set. Then, the membership function and the centers are updated until the cost function is converged. The convergence can be observed by measuring the difference in membership function changes and comparing the cluster centers at two successive iterations.

*3.2. Estimating the Population Number of Onboard, Boarding and Alighting Passengers of Each Stop using Random Forest*

Wi-Fi and Bluetooth sensing only can detect the existence of partial mobile devices within the detection range, since only the mobile devices which Wi-Fi or Bluetooth function is turned on and does not connect to any access point or devices are discoverable. Therefore, in order to estimate the population number of passengers, RF regression model was employed.

RF regression is a widely used non-parametric machine learning algorithm for classification and regression. As discussed before, RF can capture the non-linear relationship in the data set. The general concept of RF was introduced by Breiman in 2001 (Breiman, 2001). In this study, the CART (Classification and Regression Trees) algorithm (Breiman, 2017) was deployed for trees development. Once a CART tree has been built, the branches which do not contribute to the predictive performance of the tree will be pruned for avoiding overfitting. However, if the CART trees are used in the random forest, the pruning process will be ignored since the generalization error of a random forest will always converge.

For the RF model development in this study, 5 variables were selected as the regressors, including the day of week, the hour of day, the minute of hour, the dummy variable of whether the current stop is the last stop of the trip, and the number of unique MAC addresses which are clustered as passengers. The number of trees in RF would be set as $ntree = 100$, which is recommended by previous studies (Genuer et al., 2010; Oliveira et al., 2012).

*3.3. Performance Evaluation of the Proposed Methodologies*

*3.3.1. Fuzzy C-Means Clustering Performance Evaluation*

In order to evaluate the FCM clustering performance, Gaussian Mixture Model (GMM) and a Bayesian approach to GMM (BGM) were selected as the baseline models. The reason that we chose mixture density-based clustering as the baseline models is that GMM is good at forming smooth approximation to arbitrarily shaped of the probability density and at scaling with the dimensionally of data (Reynolds, 2015), which is suitable for the dataset with ambiguous boundaries. BGM optimizes the selection of the number of components in the model as well as the partition data sets by automatically penalizing the overcomplex model (Roberts et al., 1998), which could further improve the performance of the GMM model. In addition, BGM can avoid overfitting by eliminating parameters using integration (Attias, 2000). The model specification can be found in the references (Lee et al., 2003; Reynolds, 2015). Then, four measurements for clustering validation were employed. The following paragraphs would introduce them in detail.

External and internal clustering validation is the two main categories of validation methods (Wu et al., 2009). The major difference is whether external information would be used for validation. For unsupervised clustering algorithms, internal clustering validation is the only option due to the lack of available labeling information (Liu et al., 2010).

Compactness and separation are the two main criteria for measuring the validation of clusters. Compactness measures the intra-distance of each cluster and separation measures inter-distance (Tan, 2018; Zhao and Karypis, 2002). The following measurements would be employed to validate the performance of the proposed clustering algorithm, including Silhouette coefficient, Dunn's index, Davies-Bouldin index, and Beta CV measurement.

Silhouette coefficient (SC) (Rousseeuw, 1987) evaluates the performance of clustering result based on the pairwise difference of inter and intra distances of clusters, which is simply expressed as equation below

$$SC = \frac{b(j) - a(j)}{max\{a(j), b(j)\}} \quad (4)$$

where $a(j)$ is the average distance between the $i$th sample and all samples which are included in a given cluster $C_j$, and $b(j)$ is the minimum average distance between the $i$th sample and all samples of a given cluster $C_k$ ($k \neq j$). The value of SC ranges in $[-1, 1]$. A large SC value infers better clustering results.

Dunn's (DU) index is dedicated for identifying sets of compact and well separated clusters by maximizing inter-cluster distances whilst minimizing intra-cluster distances (Dunn, 1974). The Dunn's validation index is calculated as

$$DU = \min_{1 \leq i \leq c} \left\{ \min_{\substack{1 \leq i \leq C \\ j \neq i}} \left\{ \frac{\delta(C_i, C_j)}{\max_{1 \leq k \leq c}\{\Delta(X_k)\}} \right\} \right\} \quad (5)$$

where $\delta(C_i, C_j)$ measures the inter-cluster distance between $C_i$ and $C_j$, $\Delta(X_k)$ defines the intra-cluster distance of $X_k$, and $C$ is the number of clusters. A larger value of Dunn's index implies better clustering results.

Davies-Bouldin (DB) index (Davies and Bouldin, 1979) is the ratio of the sum of intra-cluster distance to inter-cluster separation, which is expressed by

$$DB = \frac{1}{k} \sum_{i=1}^{k} \max_{i \neq j} \frac{\left(D(C_i) + D(C_j)\right)}{D(v_i, v_j)} \quad (6)$$

where $D(v_i, v_j)$ is the inter-cluster distance between the centers of clusters $C_i$ and $C_j$ and $D(C_i)$ is the intra-cluster diameter of the cluster $C_i$. The lower the DB value, the better the clustering results.

Beta CV Measure (Analysis, 2011) is a measurement of clustering validation based on the ratio of the mean intra-cluster distance to the mean inter-cluster distance which can be calculated as below (Han and Science, 2017)

$$Beta\ CV = \frac{Distance_{intra}/N_{intra}}{Distance_{inter}/N_{inter}} \quad (7)$$

where $N_{intra}$ is the number of distinct intra-cluster edges, $N_{inter}$ is the number of distinct inter-cluster edges.

*3.3.2. Evaluation of the Estimated Population Number of Onboard, Boarding and Alighting Passengers of Each Stop*

The performance of the RF regression model for estimating the population number of onboard, boarding and alighting passengers of each stop is compared with the performance of the traditional linear regression model developed based on the same data set. Mean Absolute Error (MAE), Mean Square Error (MSE) and Mean Absolute Percentage Error (MAPE) are used as the evaluation measurements. The following equations present the measurement formulation,

$$MAE = \frac{\sum_{i=1}^{N}|\hat{Y}_i - Y_i|}{N} \quad (8)$$

$$MSE = \frac{\sum_{i=1}^{N}(\hat{Y}_i - Y_i)^2}{N} \quad (9)$$

$$MAPE = \frac{1}{N}\sum_{i=1}^{N}\frac{|\hat{Y}_i - Y_i|}{Y_i} \times 100\% \tag{10}$$

where $\hat{Y}_i$ is the estimated number of onboard, boarding or alighting passenger of stop $i$, $Y_i$ is the observed value, and $N$ is the number of stops in the testing data set. Typically, the MAE presents a measure of the average misprediction of the model, the MSE is used to measure the error associated with a prediction, and the MAPE usually expresses accuracy as a percentage. The model with a smaller value of MAE, MSE and MAPE performs better in the prediction of observed data.

## 4. Numerical Results

The numerical results would be presented in terms of (i) passenger and non-passenger MAC address separation, and (ii) the estimation of population number of onboard, boarding and alighting passenger of each stop.

### 4.1. Passenger and Non-Passenger MAC Address Separation

The raw Wi-Fi and Bluetooth MAC address sensing data and the GPS data were used to extract the proposed features of each MAC address. The FCM clustering was conducted to cluster each MAC address into passenger or non-passenger clusters. The measurements of each model are presented in Table 3. According to the evaluation measurements, the FCM clustering model outperformed all models in terms of achieving the highest value of SC and DU and the lowest value of Beta CV and DB, which indicates the clusters were separated well by the FCM clustering. The BGM and GM models had similar performance according to the closing value of all 4 measurements.

Table 3: Clustering algorithms' performance evaluation of extracting MAC passenger based on the extracted features

| Evaluation Measurements | Fuzzy C-means | Bayesian Gaussian Mixture | Gaussian Mixture |
|---|---|---|---|
| Silhouette Coefficient | 0.74289 | 0.65651 | 0.63654 |
| Beta CV Measure | 0.16561 | 0.21994 | 0.23426 |
| Dunn's index | 0.00021 | 0.00007 | 0.00005 |
| Davies-Bouldin Index | 0.67708 | 0.79231 | 0.81318 |

Totally, 5064 unique MAC address were clustered by the FCM clustering algorithm into two clusters with 399 passenger MAC addresses and 4665 non-passenger MAC addresses. Based on the FCM clustering results, the statistical summary of the characteristics of each proposed feature is presented in Table 4. The mean values of the number of times detected and the detection duration of passengers are much larger than those of non-passengers. The non-passengers only had 1.26 times average detection and 4.24 seconds detection duration which is consistent with the assumption that non-passengers' MAC address would be detected for a few times or within a narrow time window. The average RSSI and the max RSSI of passengers is larger than those of non-passenger for all four numbers, which is reasonable that the signal strength of non-passenger's mobile device might be influenced by the bodyshell of the transit vehicle or the larger distance with the sensor. The distances of the vehicle to the nearest station when the MAC address was first and last detected of the passenger are about 200 meters smaller than those of non-passenger for the mean values. It is explicable that the MAC addresses of passengers are more likely to be detected around the station for the first and the last detection, and MAC addresses of non-passengers are more likely to be detected during the transit vehicle is far away from stations. But the distance to the transit station of the MAC addresses of non-passengers who are waiting for other vehicles at the station is very trivial. Thus, several MAC addresses which were the first and the last detections close to the stations were still considered as non-passenger. Other three vehicle status related features of passengers, including trip distance, average speed, and max speed, have higher mean values than those of non-passengers for all four-number. The mean values of these three features of non-passengers' MAC addresses are close to zero, which indicates the vehicle almost halted during the time period when the MAC addresses of non-passengers were detected. It is noticed that the maximum values of average speed and the max speed of passenger are unreasonably high, which is caused by unstable GPS data. In summary, the feature space of the passenger's MAC address features is distinguished with the MAC address of non-passenger.

Table 4: The statistical summary of MAC address data features

| Features | Passenger | | | | Non-Passenger | | | |
|---|---|---|---|---|---|---|---|---|
| | Min | Max | Mean | S.D. | Min | Max | Mean | S.D. |
| The Number of Times Detected | 2.00 | 1021.00 | 20.89 | 69.24 | 1.00 | 31.00 | 1.26 | 1.71 |
| Detection Duration (Seconds) | 1.00 | 3060.00 | 418.78 | 679.08 | 0.00 | 1253.00 | 4.24 | 62.89 |
| Average RSSI (dBm) | -88.00 | -22.19 | -56.65 | 14.17 | -91.00 | -39.00 | -61.79 | 12.10 |
| Max RSSI (dBm) | -85.00 | -17.00 | -50.62 | 15.01 | -91.00 | -37.00 | -61.23 | 12.30 |
| Least Distance Start (Meters) | 1.64 | 2306.17 | 152.19 | 213.12 | 3.22 | 1064.48 | 324.49 | 213.49 |
| Least Distance End (Meters) | 2.18 | 1722.00 | 144.31 | 195.75 | 3.22 | 1064.48 | 325.57 | 213.74 |
| Trip Distance (Meters) | 8.94 | 20442.36 | 2409.68 | 4181.46 | 0.00 | 184.03 | 1.77 | 14.61 |
| Average Speed (Meters/Second) | 0.44 | 30.29 | 7.72 | 6.43 | 0.00 | 8.25 | 0.11 | 0.77 |
| Max Speed (Meters/Second) | 0.44 | 79.70 | 8.34 | 10.71 | 0.00 | 31.98 | 0.27 | 2.18 |

*4.2. Estimating the Population Number of Onboard, Boarding and Alighting Passenger of Each Stop*

After separating the MAC address of passengers from the whole data set, the boarding and alighting stations of each passenger need to be determined. The boarding and alighting stations of each unique MAC address were assigned based on the distances of the vehicle to the nearest station for the first and the last detection. The stations with the smallest distance to the vehicle for the first and the last detection were considered as the boarding and alighting station, respectively. The passenger was considered as staying on the vehicle for the stations between the boarding and alighting stations. The total number of onboard, boarding and alighting passengers of each stop were counted based on the FCM clustering results. In order to further evaluate the clustering results of the FCM, the total number of onboard, boarding and alighting passenger of each stop were also counted based on the BGM and GM clustering results. Then, the data were divided into training data and testing data with a portion of 7:3 for developing the proposed RF regression model as well as the linear regression model. The manual counting number of onboard, boarding and alighting passengers of each stop was used as the ground truth for calculation the evaluation measurements of MAE, MSE and MAPE.

Firstly, only the number of onboard passengers was estimated. The estimation performance evaluation results are presented in Table 5. According to the evaluation results, the estimated results, the estimated results based on the FCM clustering results performed better than other baseline models in terms of the smallest values of MSE, MAE and MAPE for both the estimations of the linear regression and the RF regression, which indicated the effectiveness of the FCM clustering model for the passenger MAC address separation. Furthermore, the estimated performance of the proposed RF regression algorithm is more accurate than that of the linear regression model for all three clustering results. MSE, MAE, and MAPE of the RF regression model are highly smaller those of the linear regression model. Then, the estimated values of each stop based on the FCM clustering results and the ground truth value were visualized in Figure 3.

Table 5: Estimates performance evaluation of the population number of passengers based on clustered number of passenger MAC addresses

| Methods | Fuzzy C-means | | | Bayesian Gaussian Mixture | | | Gaussian Mixture | | |
|---|---|---|---|---|---|---|---|---|---|
| | MSE | MAE | MAPE | MSE | MAE | MAPE | MSE | MAE | MAPE |
| Linear Regression | 20.29 | 3.26 | 28.96 | 23.54 | 3.46 | 33.86 | 27.79 | 3.43 | 34.49 |
| Random Forest | 14.61 | 2.08 | 11.27 | 22.61 | 3.25 | 31.02 | 10.36 | 2.50 | 32.09 |

The solid line in black color is the clustered number of MAC addresses belonging to the onboard passenger of each stop based on the FCM results. The solid line in red color is the ground truth number of onboard passengers of each stop. For most of the stops, the number of passenger MAC addresses is a small proportion of the ground truth, and it can effectively reveal the variation tendency of the ground truth. The dashed line in blue color presented the estimated number of onboard passengers based on the estimation of linear regression. By employing linear regression, the number of passenger MAC addresses were enlarged with a fixed proportion. It is noticed that the difference between the number of passenger MAC addresses and the ground truth is not invariable, e.g. for some stops, the difference is smaller than the general magnitude, but for some other stops, the number of passenger MAC addresses is even larger than the ground truth value. Thus, if estimating the population number of onboard passengers by regressing the number of passenger MAC addresses through a linear relationship, the errors of the estimation could be unacceptable. By capturing the non-linear relationship between the number of passenger MAC addresses and the ground truth, the RF regression model achieved a precise estimation of the population number of onboard passengers. The green dash line provided the estimation results of each stop, which is highly close to the red line and even superposed the red line for some stops.

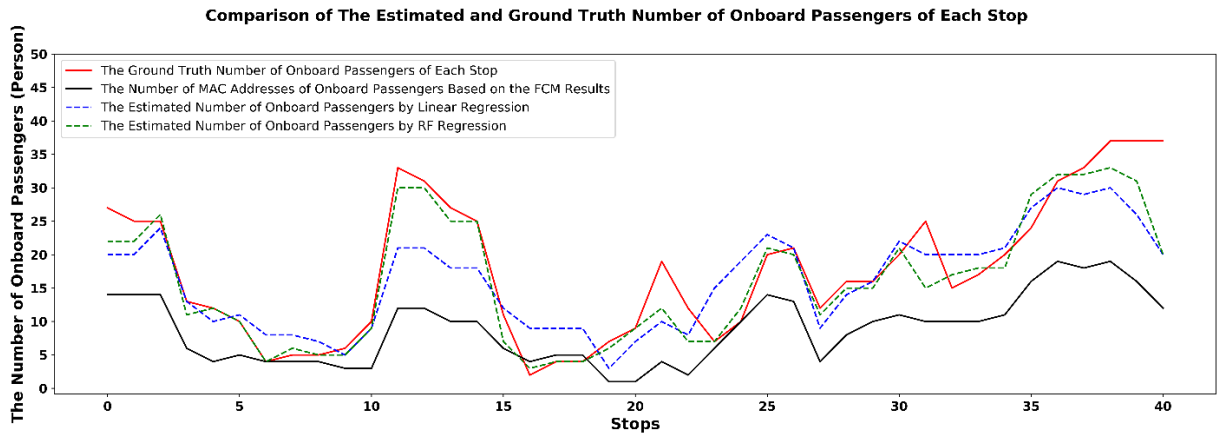

Figure 3: Comparison of clustered number of onboard passenger mac addresses based on fuzzy c-means clustering result, estimation results of the population number of onboard passengers by linear regression and random forest regression and the ground truth.

In order to further demonstrate the performance of the proposed FCM clustering algorithm, the estimated numbers of onboard passengers were calculated based on the filtered results of two existing filtering-based methods for separating out the MAC addresses of non-passengers.

Matthew et, al (2016) (Dunlap et al., 2016) developed a three-step filtering method for separating passengers and non-passengers (denotes as Filtering Method 1). The MAC address which fits any following conditions would be considered as a non-passenger MAC address, 1) total detection number is lower than 3 times for Wi-Fi MAC address and 1 time for Bluetooth MAC address, 2) the detection duration is less than 60 seconds, and 3) the distances of vehicle to the nearest station when the MAC address is detected at the first and the last time are larger than 600 ft for Wi-Fi and 300 ft for Bluetooth. The first and the last stops of the trip were determined by the stations which are the nearest stops to the vehicle when the MAC address is detected at the first and the last time. Rabi et, al. (Mishalani et al., 2016) defined a filter with four parameters to separate the passenger from noises (denotes as Filtering Method 2). The threshold they set were, 1) any unique MAC address that had duration under 3 minutes was considered as a non-passenger, 2) any unique MAC address that had maximum signal strength that is lower than 20th of the cumulative distribution of observed signal strengths was considered as a non-passenger, 3) any unique MAC address that had total travel distance less than 900 ft was considered as a non-passenger, and 4) any unique MAC address that had a total number of detected signal per mile was less than 10 was considered as a non-passenger. To determine the boarding and alighting stops for each unique MAC address of passenger, Filtering Method 2 made use of the first and last detected time of each MAC, the distance between the sensor and stops nearby as well as a preset threshold of the maximum sensor detection range of 200 ft.

The estimation results of RF regression and linear regression based on the filtering results were compared with the estimations based on FCM clustering results. Table 6 shows the evaluation measurements. Consistent with the previous evaluation results, the RF regression model performed better than the linear regression for all measurements.

Among existing filtering methods, Filtering Method 2 achieved a better performance than Filtering Method 1 by employed more filters and combining the detection ranges for boarding and alighting stop determination. All three measurements of Filtering Method 2 were smaller than those of Filtering Method 1. The estimated results based on the FCM clusters were hugely improved compared to the results of two existing filtering algorithms. It is further demonstrated that the MAC addresses of passenger and non-passenger are hard to be well separated by filters. However, by considering the overlapped feature spaces of passenger and non-passenger, the FCM clustering algorithm effectively separated the MAC addresses of passenger and non-passenger. Furthermore, the RF regression model effectively estimated the population number of onboard passengers by capturing the non-linearity.

Table 6: Performance comparison of the proposed algorithm and the existing algorithms

| Methods | Fuzzy C-means | | | Filtering Method 1 | | | Filtering Method 2 | | |
|---|---|---|---|---|---|---|---|---|---|
|  | MSE | MAE | MAPE | MSE | MAE | MAPE | MSE | MAE | MAPE |
| Linear Regression | 26.29 | 3.26 | 28.96 | 52.05 | 5.32 | 51.25 | 67.78 | 6.62 | 58.62 |
| Random Forest | 14.61 | 2.08 | 11.27 | 35.16 | 3.84 | 36.47 | 30.03 | 3.76 | 27.5 |

The scatter plots of the ground truth versus the estimated number of onboard passengers based on the FCM results and two existing filtering algorithms are presented in Figure 4 for analyzing the reason for inaccurate separation results. The RF model was used for estimating the population value based on the separated passenger MAC address counts. According to the figure, the dots in the plots of both existing filtering algorithms are dispersed around the diagonal line. For Filtering Method 1, most of the dots are above the diagonal line, which indicates the MAC addresses were more likely to be separated into the non-passenger cluster so that the number of onboard was underestimated. The potential reason is that the Filtering Method 1 is inclined to separate passenger into the non-passenger cluster, e.g. the GPS location was recorded every 20 seconds so that the distance of the vehicle to the nearest station is highly possible to be larger than the detection range for the first detection of a passenger MAC address. For the results of Filtering Method 2, the dots in the scatter plot centered more to the diagonal line. However, most of the dots are beneath the line, which indicates the algorithm overestimated the number of onboard passengers. The explanation could be the filter of signal strength was apt to separate the non-passenger MAC addresses to the passenger cluster since the distribution of signal strength of non-passenger MAC address is similar to the distribution of passenger MAC address. The rightmost scatter plot presents the data estimated based on the FCM passenger cluster data. The dots were highly concentrated around the diagonal line which is the indicator of a minor difference between the estimation and the ground truth. It is noticed that the estimation is more accurate for the small number of passengers. However, as the number increased, the error became more considerable. The potential reason could be the insufficient data point with a large value in the training data set, which can be solved by collecting more data for the training process.

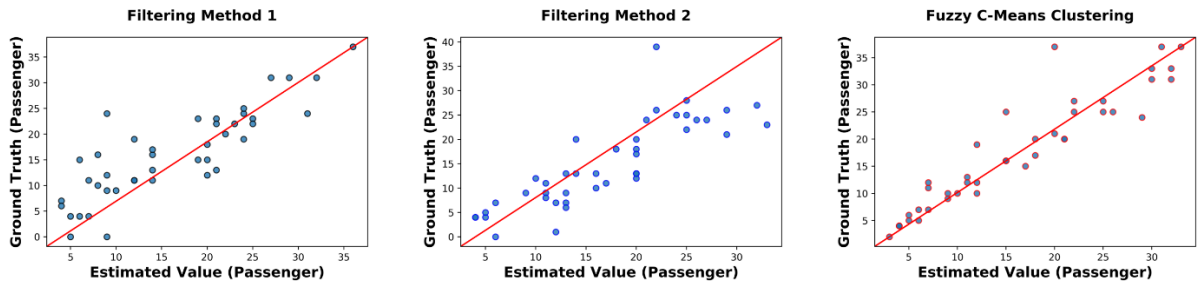

Figure 4: The scatter plot of ground truth versus estimated number of onboard passengers based on FCM clustering and two existing filtering algorithms.

Besides the number of onboard passengers, the numbers of boarding and alighting passengers of each stop were also estimated using the RF regression model based on the FCM clustering results. The estimation performance was evaluated by the three measurements calculated based on the manual counting numbers of boarding and alighting passengers of each stop, which is presented in Table 7. According to the evaluation results, the estimated numbers of boarding and alighting passengers are acceptable in terms of the small value of MSE, MAE, and MAPE. It is noticed that the MAPEs of both estimated numbers of boarding and alighting passengers are higher than the MAPE of the estimated number of onboard passengers, which is potentially caused by numerous zero values of the number of boarding and alighting passengers in the dataset.

Table 7: Estimation performance evaluation of truth number of boarding and alighting passenger of each stop

| Estimations | MSE | MAE | MAPE |
|---|---|---|---|
| Estimating the Number of Boarding Passengers of Each Stop | 0.86 | 0.50 | 14.72 |
| Estimating the Number of Alighting Passengers of Each Stop | 0.96 | 0.54 | 17.41 |

*4.3.     Estimating the Ridership Flow and O-D Information of the Selected Transit Trip*

Based on the proposed methodology, the transit demand can be monitored by the estimated numbers of onboard, boarding and alighting passengers and O-D information from Wi-Fi and Bluetooth sensing data. In order to further demonstrate the feasibility of the proposed method, the ridership flow and O-D matrix of a selected trip were estimated based on the proposed methodologies, and the results are presented in Table 8 and 9. The selected trip traveled on November 9th, 2018 from 19:38:58 to 20:05:48. Totally, the transit vehicle stopped at 15 stations during the trip. The detailed statistical description of the MAC address sensing data can be found in Table 1.

Table 8 presents the O-D matrix of the MAC addresses which were clustered in passenger clusters by the FCM clustering algorithm. Even only partial O-D information can be achieved, the main trend of the travel demand still can be achieved. Besides the O-D matrix, the numbers of boarding and alighting passengers were estimated using the RF regression model. The RF regression was trained by the data set which is collected from other trips. The ground truth numbers of boarding and alighting passengers of each stop are also presented in the table. The estimated result has an acceptable difference with the ground truth value, which demonstrates the effectiveness of the proposed method. Table 9 shows the estimated number of onboard passengers of each stop and the ground truth as well. The estimated errors are negligible for the most stops. However, the estimation errors were relatively large for the last two stops. Since the data collection sensor was powered off before the trip ended for the selected trip, the MAC address sensing data quality was influenced for the last two stops. Therefore, the zero number of MAC addresses for the last stops is the main reason for the large error. The number of passenger MAC addresses is the main input of the RF regression model for estimating the true number of passengers. The hour of the day and the minute of the hour are two additional input of the RF regression model for adjusting the estimated value. Thus, if the estimated number of passenger MAC addresses is greatly different from the ground truth, the RF regression still hardly regress the number of passenger MAC addresses close to the ground truth. This could be future work for improving the methodologies.

By successfully capturing the partial O-D matrix, the numbers of onboard, boarding and alighting passengers of each stop of the selected trip and the O-D information can be achieved. Based on the output parameters of the proposed methodologies, it can be easily observed through the table that which parts of the trip have more travel demands, which stops are more popular for the traveler, and what is the O-D trend of travelers.

**5.    REAL-TIME TRANSIT RODERSHIP FLOW AND O-D INFORMATION MONITORING SYSTEM**

Based on the proposed methods, an algorithm framework was developed for public transit passenger monitoring using Wi-Fi and Bluetooth sensing data which is presented in Figure 5. Generally, the proposed algorithm is a three-step data-driven approach. Step one aims to extract the proposed features of each MAC address and the vehicle status during the detection time period each MAC address from the Wi-Fi and Bluetooth MAC address sensing data, and high-resolution GPS location record of the transit vehicle. Then, the MAC addresses with their extracted features would be used as input data of step two. In step two, the proposed FCM clustering algorithm would be employed to cluster the MAC addresses into the passenger cluster and non-passenger cluster based on the features. The MAC addresses data which are clustered as passenger's MAC address will be the input of step three. In addition, the ground truth data collected by manual counting, in-vehicle surveillance camera, and transit smart card reader would be also used for training the proposed RF regression model. The trained RF regression model would estimate the population numbers of onboard, boarding and alighting passengers of each stop. By accurately clustering the MAC addresses of passengers, the OD matrix of partial passengers also can be achieved based on the assigned origin and destination of each passenger.

Table 8: O-D matrix of the selected trip

| Boarding \ Alighting | 1 | 2 | 3 | 4 | 5 | 6 | 7 | 8 | 9 | 10 | 11 | 12 | 13 | 14 | 15 | Total Boarding MAC | Total Ground Truth Boarding | Total Estimated Boarding |
|---|---|---|---|---|---|---|---|---|---|---|---|---|---|---|---|---|---|---|
| 1 | 0 | 0 | 0 | 0 | 0 | 0 | 0 | 0 | 0 | 0 | 0 | 0 | 0 | 2 | 0 | 2 | 2 | 3 |
| 2 |   | 0 | 0 | 0 | 0 | 0 | 0 | 0 | 0 | 0 | 0 | 0 | 0 | 0 | 1 | 1 | 2 | 0 |
| 3 |   |   | 0 | 1 | 0 | 1 | 0 | 0 | 0 | 0 | 0 | 0 | 0 | 0 | 0 | 2 | 1 | 2 |
| 4 |   |   |   | 0 | 0 | 0 | 0 | 0 | 0 | 1 | 0 | 0 | 0 | 0 | 0 | 1 | 0 | 1 |
| 5 |   |   |   |   | 0 | 1 | 0 | 0 | 0 | 0 | 0 | 0 | 0 | 0 | 0 | 1 | 1 | 1 |
| 6 |   |   |   |   |   | 0 | 1 | 0 | 0 | 0 | 0 | 0 | 0 | 0 | 0 | 1 | 1 | 1 |
| 7 |   |   |   |   |   |   | 0 | 3 | 1 | 0 | 0 | 0 | 0 | 0 | 0 | 4 | 2 | 3 |
| 8 |   |   |   |   |   |   |   | 0 | 0 | 0 | 0 | 0 | 0 | 0 | 0 | 0 | 1 | 3 |
| 9 |   |   |   |   |   |   |   |   | 0 | 0 | 0 | 0 | 2 | 0 | 0 | 2 | 0 | 2 |
| 10 |   |   |   |   |   |   |   |   |   | 0 | 1 | 1 | 0 | 0 | 0 | 2 | 2 | 1 |
| 11 |   |   |   |   |   |   |   |   |   |   | 0 | 1 | 0 | 0 | 0 | 1 | 0 | 0 |
| 12 |   |   |   |   |   |   |   |   |   |   |   | 0 | 0 | 0 | 1 | 1 | 1 | 2 |
| 13 |   |   |   |   |   |   |   |   |   |   |   |   | 0 | 1 | 0 | 1 | 0 | 3 |
| 14 |   |   |   |   |   |   |   |   |   |   |   |   |   | 0 | 0 | 0 | 3 | 2 |
| 15 |   |   |   |   |   |   |   |   |   |   |   |   |   |   | 0 | 0 | 0 | 0 |
| Total Alighting MAC | 0 | 0 | 0 | 1 | 0 | 2 | 1 | 3 | 1 | 1 | 1 | 2 | 2 | 3 | 2 | 19 / 19 |   |   |
| Total Ground Truth Alighting | 0 | 0 | 0 | 0 | 1 | 2 | 1 | 0 | 1 | 0 | 1 | 0 | 2 | 2 | 6 |   | 16 / 16 |   |
| Total Estimated Alighting | 0 | 0 | 1 | 1 | 1 | 2 | 1 | 1 | 2 | 2 | 1 | 2 | 3 | 2 | 5 |   |   | 24 / 24 |

Table 9: The number of onboard passengers of the selected trip

| Stops | 1 | 2 | 3 | 4 | 5 | 6 | 7 | 8 | 9 | 10 | 11 | 12 | 13 | 14 | 15 |
|---|---|---|---|---|---|---|---|---|---|---|---|---|---|---|---|
| Ground Truth Onboard Passenger | 2 | 4 | 5 | 5 | 5 | 4 | 5 | 6 | 5 | 7 | 6 | 7 | 5 | 6 | 6 |
| Onboard MAC of Each Stop | 2 | 3 | 5 | 5 | 6 | 5 | 8 | 5 | 6 | 7 | 7 | 6 | 5 | 2 | 0 |
| Estimated Onboard Passenger | 2 | 4 | 5 | 5 | 5 | 6 | 6 | 4 | 4 | 5 | 5 | 5 | 4 | 3 | 1 |

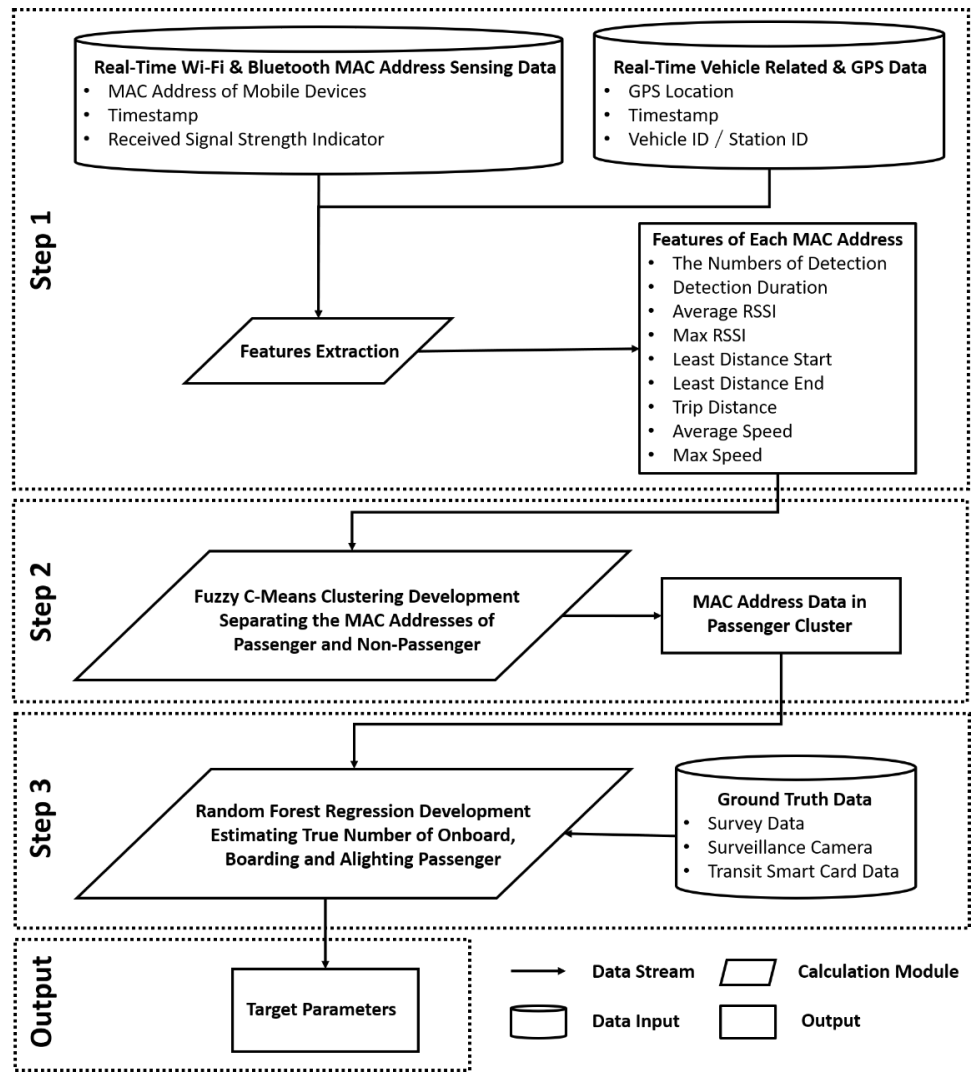

Figure 5: The proposed algorithm framework.

Finally, a dedicated system was designed for monitoring public transit passengers, that the system structure is presented in Figure 6. The data collection sensor will be installed in the vehicle as well as at transit stations. The Wi-Fi and Bluetooth MAC addresses of discoverable mobile devices within the detection range will be detected, including mobile phones, laptops, earphones, etc. The real-time data will be transmitted from sensors to the remote data management and analysis server through 3G or 4G wireless networks. In addition, transit smart card data and in-vehicle surveillance camera data will also be transmitted to the remote server for providing the ground truth information to the system. Then, the ridership flow and O-D information will be estimated through the proposed algorithm framework using both MAC address sensing data and the ground truth data. The target parameters not only include the numbers of onboard, boarding and alighting passengers of each stop, passenger OD matrix but also include the number of waiting passengers at stations and the waiting time of each passenger, with the help of the installed sensors at transit stations for monitoring waiting passengers. By deploying the proposed system, the real-time information of transit operation can be delivered to the public in time, so that the public can optimize the travel plan based on the real-time information. Furthermore, the real-time information of transit passengers also can be used to optimize the vehicle dispatching and trip scheduling.

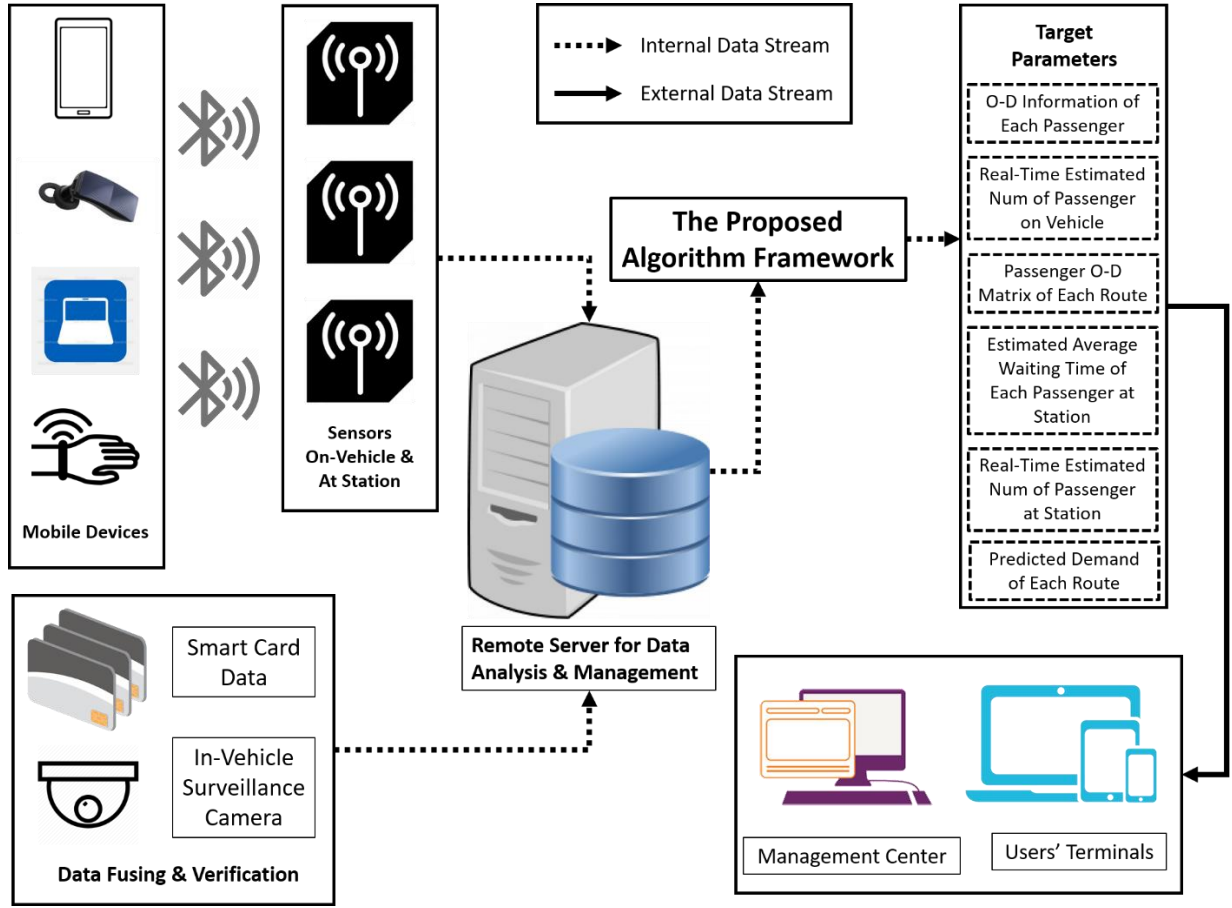

Figure 6 The structure of real-time transit passenger monitoring system.

## 6. Conclusions and Future Works

In summary, this study developed a three-step data-driven approach for mining the transit ridership flow and O-D information from Wi-Fi and Bluetooth sensing data, which is consisted of feature extraction for each detected MAC address, FCM clustering algorithm for separating MAC address of passenger based on the extracted features of each MAC address, and RF regression for estimating the population number of ridership flow based on the number of MAC addresses in passenger cluster. In order to demonstrate the effectiveness and efficiency of the proposed model, GMM and BGM were selected as the baseline models of the FCM clustering, and linear regression was selected for the RF regression. Multiple measurements were calculated based on observed data and the estimates to provide the quantitative evidence for evaluating the estimation performance. Finally, a real-time system for monitoring the ridership flow and O-D information was designed based on the proposed algorithm framework.

According to the experimental results, the proposed FCM clustering algorithm achieved the best performance compared with the baseline models. Furthermore, by well separating the passenger and non-passenger MAC addresses, the estimated population number of onboard passengers of each stop based on the FCM clustering results estimated by both linear regression and RF regression outperformed the GMM and BGM models in terms of MSE, MAE and MAPE, which further demonstrated the effectiveness of the proposed FCM clustering algorithm. In order to compare the performance of the proposed FCM model with the filter-based models in previous studies, the population number of onboard passengers was estimated based on the FCM clustering results, Filtering Method 1 and Filtering Method

2, respectively, by both linear regression and RF regression. For both two regression methods, the estimates based on the FCM clustering results performed better than others. Moreover, for all experiments of population ridership estimation conducted in this study, the proposed RF regression model outperformed the linear regression model by achieving lower estimation errors. Therefore, the proposed three-step data-driven method effectively estimated the ridership flow and O-D information from Wi-Fi and Bluetooth sensing data.

The findings of this study can help to provide precise and real-time transit ridership flow and O-D information for supporting the transit network plan and optimizing the service. In addition, transit service users could get a better understanding of the operational status of the transit system for saving the waiting time and avoiding congestion. In this study, the O-D information is still a part of the population O-D information. The O-D population inference based on Wi-Fi and Bluetooth sensing data could be a valuable research direction for improving the comprehensiveness and accuracy of the whole system.

**Acknowledgements**

This research was supported by the multi-institutional project (Exploring Weather-related Connected Vehicle Applications for Improved Winter Travel in Pacific Northwest) of Pacific Northwest Transportation Consortium (PacTrans) USDOT University Transportation Center for Federal Region 10.

**Reference**

Global mobile phone internet user penetration 2019 | Statistic [WWW Document], n.d. URL https://www.statista.com/statistics/284202/mobile-phone-internet-user-penetration-worldwide/ (accessed 3.13.19).

Smartphone penetration in the US (share of population) 2010-2021 | Statistic [WWW Document], n.d. URL https://www.statista.com/statistics/201183/forecast-of-smartphone-penetration-in-the-us/ (accessed 3.13.19).

Analysis, D., 2011. Clustering Quality Assessment.

Attias, H., 2000. A variational baysian framework for graphical models, in: Advances in Neural Information Processing Systems. pp. 209–215.

Ben-Akiva, M., Macke, P.P., Hsu, P.S., 1985. Alternative methods to estimate route-level trip tables and expand on-board surveys.

Bezdek, J.C., Ehrlich, R., Full, W., 1984. FCM: The fuzzy c-means clustering algorithm. Comput. Geosci. 10, 191–203.

Breiman, L., 2017. Classification and regression trees. Routledge.

Breiman, L., 2001. Random forests. Mach. Learn. 45, 5–32.

Cunche, M., 2014. I know your MAC Address: Targeted tracking of individual using Wi-Fi. J. Comput. Virol. Hacking Tech. 10, 219–227.

Davies, D.L., Bouldin, D.W., 1979. A cluster separation measure. IEEE Trans. Pattern Anal. Mach. Intell. 224–227.

Duives, D.C., Daamen, W., Hoogendoorn, S.P., 2018. How to Measure Static Crowds? Monitoring the Number of Pedestrians at Large Open Areas by Means of Wi-Fi Sensors.

Dunlap, M., Li, Z., Henrickson, K., Wang, Y., 2016. Estimation of origin and destination information from Bluetooth and Wi-Fi sensing for transit. Transp. Res. Rec. J. Transp. Res. Board 11–17.

Dunn, J.C., 1974. Well-separated clusters and optimal fuzzy partitions. J. Cybern. 4, 95–104.

Gan, S., Liang, S., Li, K., Deng, J., Cheng, T., 2018. Trajectory length prediction for intelligent traffic signaling: A data-driven approach. IEEE Trans. Intell. Transp. Syst. 19, 426–435.

Gastaldi, M., Gecchele, G., Rossi, R., 2014. Estimation of Annual Average Daily Traffic from one-week traffic counts. A combined ANN-Fuzzy approach. Transp. Res. Part C Emerg. Technol. 47, 86–99.

Genuer, R., Poggi, J.-M., Tuleau-Malot, C., 2010. Variable selection using random forests. Pattern Recognit. Lett. 31, 2225–2236.

Han, J., Science, C., 2017. CS 412 Intro . to Data Mining.

Hathaway, R.J., Bezdek, J.C., 2001. Fuzzy c-means clustering of incomplete data. IEEE Trans. Syst. Man, Cybern. Part B 31, 735–744.

Hidayat, A., Terabe, S., Yaginuma, H., 2018. WiFi Scanner Technologies for Obtaining Travel Data about Circulator Bus Passengers: Case Study in Obuse, Nagano Prefecture, Japan. Transp. Res. Rec. 0361198118776153.

Ji, Y., Zhao, J., Zhang, Z., Du, Y., 2017. Estimating bus loads and OD flows using location-stamped farebox and Wi-Fi signal data. J. Adv. Transp. 2017.

Kieu, L.-M., Bhaskar, A., Chung, E., 2015. A modified Density-Based Scanning Algorithm with Noise for spatial travel pattern analysis from Smart Card AFC data. Transp. Res. Part C Emerg. Technol. 58, 193–207.

Kostakos, V., Camacho, T., Mantero, C., 2013. Towards proximity-based passenger sensing on public transport buses. Pers. ubiquitous Comput.

17, 1807–1816.

Kostakos, V., Camacho, T., Mantero, C., 2010. Wireless detection of end-to-end passenger trips on public transport buses, in: 13th International IEEE Conference on Intelligent Transportation Systems. pp. 1795–1800.

Kusakabe, T., Asakura, Y., 2014. Behavioural data mining of transit smart card data: A data fusion approach. Transp. Res. Part C Emerg. Technol. 46, 179–191.

Lee, D.-S., Hull, J.J., Erol, B., 2003. A Bayesian framework for Gaussian mixture background modeling, in: Proceedings 2003 International Conference on Image Processing (Cat. No. 03CH37429). pp. III--973.

Lesani, A., Miranda-Moreno, L.F., 2016. Development and Testing of a Real-Time WiFi-Bluetooth System for Pedestrian Network Monitoring and Data Extrapolation.

Liu, Y., Li, Z., Xiong, H., Gao, X., Wu, J., 2010. Understanding of internal clustering validation measures, in: 2010 IEEE International Conference on Data Mining. pp. 911–916.

Ma, X., Liu, C., Wen, H., Wang, Y., Wu, Y.-J., 2017. Understanding commuting patterns using transit smart card data. J. Transp. Geogr. 58, 135–145.

Ma, X., Wang, Y., Chen, F., Liu, J., 2012. Transit smart card data mining for passenger origin information extraction. J. Zhejiang Univ. Sci. C 13, 750–760.

Ma, X., Wu, Y.-J., Wang, Y., Chen, F., Liu, J., 2013. Mining smart card data for transit riders' travel patterns. Transp. Res. Part C Emerg. Technol. 36, 1–12.

Malinovskiy, Y., Wu, Y.-J., Wang, Y., Lee, U.K., 2010. Field experiments on bluetooth-based travel time data collection.

Mishalani, R.G., McCord, M.R., Reinhold, T., 2016. Use of Mobile Device Wireless Signals to Determine Transit Route-Level Passenger Origin--Destination Flows: Methodology and Empirical Evaluation. Transp. Res. Rec. J. Transp. Res. Board 123–130.

Morency, C., Trépanier, M., Agard, B., 2007. Measuring transit use variability with smart-card data. Transp. Policy 14, 193–203.

Oliveira, S., Oehler, F., San-Miguel-Ayanz, J., Camia, A., Pereira, J.M.C., 2012. Modeling spatial patterns of fire occurrence in Mediterranean Europe using Multiple Regression and Random Forest. For. Ecol. Manage. 275, 117–129.

Oransirikul, T., Nishide, R., Piumarta, I., Takada, H., 2014. Measuring bus passenger load by monitoring wi-fi transmissions from mobile devices. Procedia Technol. 18, 120–125.

Reynolds, D., 2015. Gaussian mixture models. Encycl. biometrics 827–832.

Roberts, S.J., Husmeier, D., Rezek, I., Penny, W., 1998. Bayesian approaches to Gaussian mixture modeling. IEEE Trans. Pattern Anal. Mach. Intell. 20, 1133–1142.

Rousseeuw, P.J., 1987. Silhouettes: a graphical aid to the interpretation and validation of cluster analysis. J. Comput. Appl. Math. 20, 53–65.

Tan, P.-N., 2018. Introduction to data mining. Pearson Education India.

Upton, E., Halfacree, G., 2014. Raspberry Pi user guide. John Wiley & Sons.

Wardman, M., 1988. A comparison of revealed preference and stated preference models of travel behaviour. J. Transp. Econ. policy 22, 71–91.

Wu, J., Xiong, H., Chen, J., 2009. Adapting the right measures for k-means clustering, in: Proceedings of the 15th ACM SIGKDD International Conference on Knowledge Discovery and Data Mining. pp. 877–886.

Xu, R., Wunsch, D.C., 2005. Survey of clustering algorithms.

Zadeh, L.A., 1965. Fuzzy sets. Inf. Control 8, 338–353.

Zhao, Y., Karypis, G., 2002. Evaluation of hierarchical clustering algorithms for document datasets, in: Proceedings of the Eleventh International Conference on Information and Knowledge Management. pp. 515–524.